\documentclass[aps,preprintnumbers,amsmath,amssymb,prd,nofootinbib, preprint]{revtex4-1}
\pdfoutput=1

\usepackage{graphicx}
\usepackage{epstopdf}
\usepackage{dcolumn}
\usepackage{bm}
\usepackage{hyperref}
\usepackage{color}
\usepackage{amsmath}

\numberwithin{equation}{section}

\makeatletter
\renewcommand{\p@subsection}{}
\renewcommand{\p@subsubsection}{}
\makeatother

\begin{document}


\def\a{\alpha}
\def\b{\beta}
\def\c{\varepsilon}
\def\d{\delta}
\def\e{\epsilon}
\def\f{\phi}
\def\g{\gamma}
\def\h{\theta}
\def\k{\kappa}
\def\l{\lambda}
\def\m{\mu}
\def\n{\nu}
\def\p{\psi}
\def\q{\partial}
\def\r{\rho}
\def\s{\sigma}
\def\t{\tau}
\def\u{\upsilon}
\def\v{\varphi}
\def\w{\omega}
\def\x{\xi}
\def\y{\eta}
\def\z{\zeta}
\def\D{\Delta}
\def\G{\Gamma}
\def\H{\Theta}
\def\L{\Lambda}
\def\F{\Phi}
\def\P{\Psi}
\def\S{\Sigma}

\def\o{\over}
\def\beq{\begin{align}}
\def\eeq{\end{align}}
\newcommand{\gsim}{ \mathop{}_{\textstyle \sim}^{\textstyle >} }
\newcommand{\lsim}{ \mathop{}_{\textstyle \sim}^{\textstyle <} }
\newcommand{\vev}[1]{ \left\langle {#1} \right\rangle }
\newcommand{\bra}[1]{ \langle {#1} | }
\newcommand{\ket}[1]{ | {#1} \rangle }
\newcommand{\EV}{ {\rm eV} }
\newcommand{\KEV}{ {\rm keV} }
\newcommand{\MEV}{ {\rm MeV} }
\newcommand{\GEV}{ {\rm GeV} }
\newcommand{\TEV}{ {\rm TeV} }
\newcommand{\1}{\mbox{1}\hspace{-0.25em}\mbox{l}}
\newcommand{\headline}[1]{\noindent{\bf #1}}
\def\diag{\mathop{\rm diag}\nolimits}
\def\Spin{\mathop{\rm Spin}}
\def\SO{\mathop{\rm SO}}
\def\O{\mathop{\rm O}}
\def\SU{\mathop{\rm SU}}
\def\U{\mathop{\rm U}}
\def\Sp{\mathop{\rm Sp}}
\def\SL{\mathop{\rm SL}}
\def\tr{\mathop{\rm tr}}
\def\mpl{M_{\rm Pl}}

\def\IJMP{Int.~J.~Mod.~Phys. }
\def\MPL{Mod.~Phys.~Lett. }
\def\NP{Nucl.~Phys. }
\def\PL{Phys.~Lett. }
\def\PR{Phys.~Rev. }
\def\PRL{Phys.~Rev.~Lett. }
\def\PTP{Prog.~Theor.~Phys. }
\def\ZP{Z.~Phys. }

\def\dd{\mathrm{d}}
\def\ff{\mathrm{f}}
\def\BH{{\rm BH}}
\def\inf{{\rm inf}}
\def\ev{{\rm evap}}
\def\eq{{\rm eq}}
\def\SM{{\rm sm}}
\def\Mpl{M_{\rm Pl}}
\def\GeV{{\rm GeV}}
\newcommand{\Red}[1]{\textcolor{red}{#1}}
\newcommand{\TL}[1]{\textcolor{blue}{\bf TL: #1}}



\title{
Gravitino Production Suppressed by Dynamics of Sgoldstino
}

\author{Raymond T. Co}
\author{Keisuke Harigaya}

\affiliation{Department of Physics, University of California, Berkeley, California 94720, USA}
\affiliation{Theoretical Physics Group, Lawrence Berkeley National Laboratory, Berkeley, California 94720, USA}

\begin{abstract}
In supersymmetric theories, the gravitino is abundantly produced in the early Universe from thermal scattering, resulting in a strong upper bound on the reheat temperature after inflation.
We point out that the gravitino problem may be absent or very mild due to the early dynamics of a supersymmetry breaking field,~i.e.~a sgoldstino. 
In models of low scale mediation, the field value of the sgoldstino determines the mediation scale
and is in general different in the early Universe from the present one.
A large initial field value since the era of the inflationary reheating suppresses the gravitino production significantly.
We investigate in detail the cosmological evolution of the sgoldstino and show that the reheat temperature may be much higher than the conventional upper bound,
restoring the compatibility with thermal leptogenesis.
\end{abstract}

\date{\today}

\maketitle

\tableofcontents

\section{Introduction}

One of the most challenging puzzles in the standard model is the hierarchy problem, in which the Higgs mass is unstable against quantum corrections at high energy scales. As one of the most motivated solutions, supersymmetry (SUSY) ensures the cancellation of quantum corrections between the SM particles and their superpartners, which considerably relaxes the hierarchy problem~\cite{MaianiLecture,Veltman:1980mj,Witten:1981nf,Kaul:1981wp}.
On the other hand, gauge coupling unification at a high energy scale gives strong hints to the Grand Unified Theories (GUTs). Remarkably, supersymmetric GUTs do not suffer from the proton decay problem faced by the standard model GUTs, and further improve the precision of gauge coupling unification~\cite{Ellis:1990wk, Amaldi:1991cn, Giunti:1991ta}. 

Despite all the successes in particle physics, supersymmetry is known to create cosmological difficulties. In the case of low scale mediation of supersymmetry breaking such as gauge mediation, the gravitino is much lighter than the weak scale and is often the lightest supersymmetric particle. The gravitino is abundantly produced from the scattering of the thermalized particles in the early Universe~\cite{Pagels:1981ke,Weinberg:1982zq,Khlopov:1984pf,Ellis:1984er}.
In order not to overproduce gravitino dark matter,
the reheat temperature after inflation $T_{R}$ must be sufficiently low, $T_{R} \lsim 10^6~{\rm GeV}(m_{3/2}/{\rm GeV})$~\cite{Moroi:1993mb}, where $m_{3/2}$ is the gravitino mass,
and this bound strongly restricts the cosmological history including inflation models and baryogenesis. Especially, $T_{\rm R} < 10^9$ GeV is in conflict with thermal leptogenesis~\cite{Fukugita:1986hr}. This is known as the gravitino problem in low scale mediation of supersymmetry breaking.

Several solutions have been considered so far.
One may assume a non-conventional cosmology model with a large amount of dilution from the decay of a long-lived particle~\cite{Ibe:2006rc,Hamaguchi:2009hy,Fukushima:2012ra,Kim:1992eu,Lyth:1993zw,Kawasaki:2008jc,Co:2016fln,Co:2017orl,Fujii:2002fv,Ibe:2010ym}. The large entropy production needed to reproduce the observed dark matter abundance also dilutes away the baryon asymmetry created previously,
which calls for an efficient mechanism of baryogenesis. For example, the observed baryon asymmetry can be explained by thermal leptogenesis only if the reheat temperature is extremely high, $T_R\gsim 10^{16}~{\rm GeV} (m_{3/2} / {\rm GeV})$.
Refs.~\cite{Choi:1999xm,Fukushima:2013vxa} introduce a low messenger scale and a small coupling between the goldstino component of the gravitino and the messenger.
The gravitino production is then suppressed at a temperature higher than the messenger scale.
The suppressed production helps reduce the dilution factor needed  and thus relaxes the stringent lower bound on the reheat temperature from thermal leptogenesis.
A different solution in Ref.~\cite{Badziak:2015dyc} involves an additional field whose field value determines the coupling between the messenger and the goldstino.
By a smaller field value and thus a smaller coupling in the early Universe, the upper bound on the reheat temperature is relaxed.

The interaction rate between the thermal bath and the gravitino is suppressed by the mediation scale, which is given by the field value of the scalar component of the SUSY breaking field (sgoldstino).
We point out that if the sgoldstino potential is flat enough, the field value may be large in the early Universe, suppressing the gravitino production.
We study the dynamics of the sgoldstino including thermal effects, and find that the reheat temperature may be much higher than the conventional upper bound.
The compatibility of our scenario with thermal leptogenesis is also investigated.
We emphasize that this suppression mechanism is a result of a thorough analysis of the dynamics of the existing fields necessary for low scale mediation, and can be applicable to a broad class of models with a sufficiently flat sgoldstino potential.

\section{Review of the Gravitino Problem in Gauge Mediation}

We first review gauge mediation and the production of gravitinos from the thermal bath. The SUSY breaking field $S$ is coupled to the messenger field $Q$ and $\bar{Q}$ via the superpotential term
\begin{equation}
W = y S Q \bar{Q},
\label{eq:WSQQ}
\end{equation}
which in turn generates the following term in the Lagrangian when $Q$ is integrated out
\begin{equation}
\mathcal{L} = \sum_i \int d\theta^2 \frac{\alpha_i}{4\pi} \frac{S}{v_S} W_i^\alpha W_{i \, \alpha},
\end{equation}
where $i$ is summed over $(U(1), SU(2), SU(3))$ and $v_S$ is the vev of the scalar component of $S$.
Here we assume that $Q$ and $\bar{Q}$ form a complete multiplet of $SU(5)$ GUT.
We parametrize the $F$ term of $S$ as
\begin{equation}
F_S = k \sqrt{3} m_{3/2} M_{Pl}  ,
\end{equation}
where $k \le 1$ parametrizes the fractional contribution to SUSY breaking and $M_{Pl} = 2.4 \times 10^{18}$ GeV is the reduced Planck mass. The gaugino mass is then given by
\begin{equation}
m_i = \frac{\alpha_i}{4\pi} \frac{F_S}{v_S} = \frac{\sqrt{3} \alpha_i}{4\pi} \frac{k m_{3/2} M_{Pl}}{v_S} .
\label{eq:gauginoM}
\end{equation}

The viable parameter space is as follows. To prevent $Q$ from being tachyonic, we require 
\begin{equation}
y \ge \left( \frac{4\pi}{\alpha_i} \right)^2 \frac{m_i^2}{k \sqrt{3} m_{3/2} M_{Pl}},
\label{eq:ymin}
\end{equation}
while to ensure that the quantum corrections to the $S$ mass do not exceed its vacuum mass
\begin{equation}
\Delta m_S^2 = \frac{y^2}{16\pi^2}  \frac{k^2 m_{3/2}^2 M_{Pl}^2}{v_S^2} < m_S^2,
\end{equation}
we impose the condition that 
\begin{equation}
y \le \alpha_i \frac{m_S}{m_i}  .
\label{eq:ymax}
\end{equation}
The consistency between Eqs.~(\ref{eq:ymin}) and (\ref{eq:ymax}) holds only if 
\begin{equation}
k > \frac{16\pi^2}{\alpha_i^3} \frac{m_i^3}{\sqrt{3} m_{3/2} M_{Pl} m_S} .
\label{eq:kmin}
\end{equation}
In the class of models where the low energy effective superpotential of $S$ is given by $W \simeq \sqrt{3} k m_{3/2} M_{Pl}S$,%
\footnote{This is not the case, for example, in a model of indirect gauge mediation with a superpotential $W = \lambda S \phi_1 \phi_2$ and the fields $\phi_1$ and $\phi_2$ obtain negative soft masses by a coupling with a SUSY breaking sector. The masses of $\phi_1$ and $\phi_2$ are as large as that of $S$, and we may not integrate them out.}
the supergravity effect generates a tadpole term of $S$, $V(S) = - \sqrt{3} k m_{3/2}^2 M_{Pl} S$.
The tadpole term places a minimum on the vev today, $v_S$, unless the vev is fine tuned. This translates into a lower bound on the $S$ mass
\begin{equation}
m_S \gsim 10 \left( m_3 m_{3/2} \right)^{ \scalebox{1.01}{$\frac{1}{2}$} } \simeq 300 \, \GeV \left( \frac{m_3}{\rm TeV} \right)^{ \scalebox{1.01}{$\frac{1}{2}$} } \left( \frac{m_{3/2}}{\GeV} \right)^{ \scalebox{1.01}{$\frac{1}{2}$} }.
\label{eq:mSmin}
\end{equation}

The functional form of the gravitino abundance produced  at a temperature $T$ is derived as follows;
\begin{equation}
\frac{\rho_{3/2}}{s} \simeq \frac{m_{3/2} n_i^2 \sigma_i v}{H s} \simeq \frac{m_{3/2} k^2 M_{Pl}}{v_S^2} T \simeq \left(\frac{4\pi}{\alpha_i} \right)^2 \frac{m_i^2}{3 m_{3/2}M_{Pl}} T ,
\label{eq:rho32s}
\end{equation}
where $\rho$ and $s$ are the energy and entropy density respectively, while $\sigma_i$ refers to the scattering cross section between the gravitino and the gaugino/gauge boson, which follows the thermal equilibrium number density $n_i$.
Here we assume that the temperature is sufficiently small so that the gravitino is not thermalized.
As can be seen in Eq.~(\ref{eq:rho32s}), the production mode by thermal scattering is dominated at higher temperature, which we call ``UV dominated," and peaked at the reheat temperature after inflation $T_R$.
The precise result of Eq.~(\ref{eq:rho32s}) is derived e.g.~in Refs.~\cite{Moroi:1993mb, Rychkov:2007uq,Kawasaki:2008qe}, which translates into the constraint on $T_R$
\begin{equation}
T_R \le 5 \times 10^6 \ \GeV \left( \frac{m_{3/2}}{\GeV} \right) \left( \frac{\rm TeV}{m_3} \right)^2 \equiv T_{\rm co}.
\label{eq:TRlimit}
\end{equation}
For $m_{3/2} \lsim 1$ MeV the upper bound is smaller than the typical gaugino mass, invalidating Eq.~(\ref{eq:TRlimit}).

The spin-3/2 component of the gravitino is also produced from the thermal bath via Planck-scale suppressed interactions.
Using the result in Ref.~\cite{Kawasaki:2008qe}, we obtain an upper bound on $T_R$,
\begin{equation}
T_R \le 2\times 10^{12}~{\rm GeV} \left( \frac{{\rm GeV}}{m_{3/2}} \right).
\label{eq:TRspin32}
\end{equation}
Although the constraint is much weaker than the one in Eq.~(\ref{eq:TRlimit}), it will be important in our mechanism where the production of the spin-1/2 component is suppressed.

\section{Sgoldstino Dynamics as a Solution}

We propose a new cosmological scenario of gauge mediation where the gravitino problem is much milder. In Eq.~(\ref{eq:rho32s}), it is assumed that $v_S$ has been a constant from the inflationary reheating until today. This is, however, not necessarily the case.
In this section, we explore the possibility that the field value $v_S(T)$ of the sgoldstino evolves with the temperature according to its potential energy $V(S)$. In particular, we consider the case where the initial field value of the sgoldstino, $v_{S0}$, is much larger than today's vev $v_S$. Based on Eq.~(\ref{eq:rho32s}), a large initial field value results in the suppression of the gravitino interaction with the thermal bath in the early Universe. 
We refer readers to Ref.~\cite{Mukaida:2012qn} and the references therein for discussions on the evolution of a scalar field in the early Universe including thermal effects.

We can parametrize the temperature dependence of the sgoldstino oscillation amplitude as $v_S(T) \propto T^n$. It is striking that the gravitino production from thermal scattering given in Eq.~(\ref{eq:rho32s}) is dominated at a lower temperature, which we call ``IR dominated," for any $n>1/2$, which is easily satisfied by the typical polynomial and logarithmic potentials. As a result, the gravitino production is insensitive to the reheat temperature.
In the case with no dilution from entropy production, the conventional constraint on the reheat temperature, $T_{\rm co}$, can be evaded as long as the combination $T/v_S^2(T)$ in Eq.~(\ref{eq:rho32s}) never exceeds $T_{\rm co}/v_S^2$ for any $T$. In general, the constraint with dilution is
\begin{equation}
{\rm max}\left(\frac{T}{v_S^2(T)} \right) \frac{1}{D} \leq \frac{T_{\rm co}}{v_S^2} ,
\label{eq:TRcons}
\end{equation}
where max($f(T)$) refers to the maximum value of $f(T)$ throughout the cosmological evolution.
This is more likely the case for quadratic and logarithmic potentials because steeper potentials lead to a smaller initial field value of $S$ as well as an earlier onset of the oscillation.

\subsection{Evolution of the Sgoldstino Field}
\label{sec:SgDyn}

We first consider the case where the sgoldstino field begins to oscillate via thermal effects.
Through the coupling with $S$ in Eq.~(\ref{eq:WSQQ}), $Q$ obtains a large mass from the large field value of the sgoldstino and further generates the thermal logarithmic potential for $S$
\begin{equation}
V_{\rm th}(S) = a_0 \, \alpha_3(T)^2 T^4 \ln \left(\frac{y^2 S^2}{T^2} \right) ,
\label{eq:thLog}
\end{equation}
where $a_0$ is a constant of order unity \cite{Anisimov:2000wx} and the logarithmic temperature dependence of $\alpha_3(T)$ will be neglected for simplicity. Here it is assumed that the messenger mass is larger than the temperature and we verify that this is true in the entire allowed parameter space.

The condition for the onset of the oscillations during inflationary reheating is given by $V_{\rm th}'' (v_{S0}) \gsim H^2$, which leads to
\begin{equation}
\alpha_3 \frac{T^2}{v_{S0}} \gsim \sqrt{\frac{\pi^2 g_*}{90}} \frac{T^4}{T_R^2 M_{Pl}},
\end{equation}
where $g_*$ is the effective number of relativistic species.%
\footnote{Here it is assumed that the radiation produced by the decay of the inflaton is thermalized and follows thermodynamics. See~\cite{Harigaya:2013vwa} and the references therein for discussion on the thermalization process.}
The oscillation temperature reads
\begin{equation}
T_{\rm osc} \simeq T_R \left( \frac{90}{\pi^2 g_*} \right)^{ \scalebox{1.01}{$\frac{1}{4}$} } \sqrt{ \frac{\alpha_3 M_{Pl}}{v_{S0}} } .
\end{equation}
We define  
\begin{equation}
\delta \equiv  \left(\frac{\pi^2 g_*}{90}\right)^{ \scalebox{1.01}{$\frac{3}{8}$} } \frac{v_{S0}}{\alpha_3 M_{Pl}} 
\label{eq:deltaDef}
\end{equation}
to parametrize the initial field value and this particular definition of $\delta$ simplifies the numerical pre-factors in the following derivations.
Here it is implicitly assumed that
\begin{equation}
\label{eq:condition_thermal}
v_{S0} \lsim \alpha_3 M_{Pl} \sqrt{\frac{90}{\pi^2g_*}}~{\rm \ \ \ \ \ \ \ and \ \ \ \ \ \ \  }~T_R > \sqrt{ \frac{m_S v_{S0}}{\alpha_3} },
\end{equation}
so that the sgoldstino begins its oscillation by the thermal logarithmic potential. If 
one of these conditions is violated, the sgoldstino begins its oscillation via its temperature independent potential.
The evolution of the sgoldstino for that case is discussed later.

The amplitude of the oscillation, $v_S(T)$, evolves as follows. The mass of $S$ is given by $\alpha_3 T^2 / v_S(T)$. Then the number density of $S$ is proportional to $T^2 v_S(T)$, which decreases with $a^{-3}$.
During the inflaton dominated era and the radiation dominated era $a^{-3} \propto T^8$ and $T^3$, and hence $v_S(T) \propto T^6$ and $T$, respectively.
The field value of the sgoldstino at the reheat temperature is then given by
\begin{equation}
\label{eq:vev_TR}
v_{S}(T_R) = v_{S0} \left( \frac{T_R}{T_{\rm osc}} \right)^6 \simeq \delta^4 \alpha_3 M_{Pl}  .
\end{equation}
After reheating, the field value evolves as
\begin{equation}
\label{eq:VSTlog}
v_{S}(T) = v_S(T_R) \frac{T}{T_R}  \simeq  \delta^4 \alpha_3 M_{Pl}  \frac{T}{T_R}.
\end{equation}

In the above analysis, we assume that reheating is caused by a perturbative decay of the inflaton. It is also possible that the reheating is caused by other dynamics such as the scattering with the thermal bath. In this case the relation between the initial field value of the sgoldstino and the field value at $T_R$ is different from Eq.~(\ref{eq:vev_TR}). It is also possible that a large Hubble induced mass term of the sgoldstino causes non-trivial dynamics of the sgoldstino before the completion of reheating. For those cases, one may still use $\delta^4$ to parametrize $v_{S}(T_R)$ without changing the discussion below.

As the temperature drops, the thermal logarithmic potential in Eq.~(\ref{eq:thLog}) becomes less effective and eventually becomes subdominant to the vacuum potential.
To be concrete, we assume that the vacuum potential is given by a simple quadratic one,  
\begin{equation}
V_{\rm vac}(S) = m_S^2 |S- v_S|^2~~(\simeq m_S^2 |S|^2~{\rm for}~v_S(T) \gg v_S).
\label{eq:VvacS}
\end{equation}
The transition to the quadratic potential occurs at the temperature $T_2$ defined by $V_{\rm th}(v_S(T_2)) = V_{\rm vac}(v_S(T_2))$,
\begin{equation}
T_2 \simeq \delta^4 \frac{m_S M_{Pl}}{T_R} .
\end{equation}
Note that $T_2 < T_R$ as long as the conditions in Eq.~(\ref{eq:condition_thermal}) are satisfied.
We now quantify $v_S(T_2)$ in relation to $v_S$. This will tell us whether the gravitino production actually becomes enhanced instead by $v_S(T) < v_S$ because the sgoldstino oscillates around the minimum at the origin set by $V_{\rm th}(v_S(T))$. 
\begin{align}
\label{eq:vST2vS}
\frac{v_S(T_2)}{v_S} = & \, \frac{v_S(T_R)}{v_S} \frac{T_2}{T_R} = \delta^8 \frac{4\pi}{\sqrt{3}} \frac{M_{Pl} m_S m_3}{k T_R^2 m_{3/2}} \nonumber \\
\simeq & \, 5 \frac{\delta^8}{k} \left( \frac{10^{12} \, \GeV}{T_R} \right)^2 \left(\frac{m_S}{300 \, \GeV}\right) \left( \frac{m_3}{\rm TeV} \right)^{ \scalebox{1.01}{$\frac{1}{2}$} } \left( \frac{\GeV}{m_{3/2}} \right).
\end{align}
When this ratio is larger than unity, which is the case for the most of the allowed parameter space, before $v_S(T)$ drops to $v_S$, $S$ starts to follow $V_{\rm vac}(S)$ and oscillates around the minimum today $v_S$. After $T_2$, $v_S(T)$ continues to decrease as $T^{3/2}$ until the temperature $T_S$, at which the amplitude is as large as the vev $v_S$. Using $v_S(T_S) = v_S$, one obtains
\begin{equation}
T_S = T_2 \left( \frac{v_S}{v_S(T_2)} \right)^{ \scalebox{1.01}{$\frac{2}{3}$} } \simeq 2 \times 10^8 \, \GeV  \left( \frac{k}{\delta^2} \right)^{ \scalebox{1.01}{$\frac{2}{3}$} } \left( \frac{T_R}{10^{12} \, \GeV} \right)^{ \scalebox{1.01}{$\frac{1}{3}$} } \left(\frac{m_S}{300 \, \GeV}\right)^{ \scalebox{1.01}{$\frac{1}{3}$} } \left( \frac{\rm TeV}{m_3} \right)^{ \scalebox{1.01}{$\frac{2}{3}$} } \left( \frac{m_{3/2}}{\GeV} \right)^{ \scalebox{1.01}{$\frac{2}{3}$} } .
\end{equation}
When the ratio is smaller than unity, $v_S (T)$ drops below $v_S$. After $T_2$, $S$ follows $V_{\rm vac}(S)$ and oscillates around the minimum today $v_S$. After the few oscillations by $V_{\rm vac}(S)$, $v_S(T)$ increases and quickly becomes as large as $v_S$.

When the initial field value of the sgoldstino is large or the reheat temperature is small, the sgoldstino begins its oscillation by the quadratic potential $V_{\rm vac}(S)$, rather than the thermal potential $V_{\rm th}(S)$. This occurs if the condition in Eq.~(\ref{eq:condition_thermal}) is violated, namely,
\begin{equation}
v_{S0} > \alpha_3 M_{Pl} \sqrt{\frac{90}{\pi^2g_*}}~{\rm \ \ \ \ \ \ \ or \ \ \ \ \ \ \  }~T_R <  \sqrt{ \frac{m_S v_{S0}}{\alpha_3} }.
\end{equation}
The oscillation temperature becomes independent of the initial amplitude and reads
\begin{equation}
T_{\rm osc} = \left( \frac{90}{\pi^2 g_*} \right)^{ \scalebox{1.01}{$\frac{1}{8}$} } \left( m_S M_{Pl} T_R^2 \right)^{ \scalebox{1.01}{$\frac{1}{4}$} },
\end{equation}
where $T_{\rm osc} > T_R$ is assumed.
The field value of the sgoldstino at the reheat temperature is given by
\begin{equation}
v_{S}(T_R) = \delta \left( \frac{\pi^2 g_*}{90}\right)^{ \scalebox{1.01}{$\frac{1}{8}$} } \frac{\alpha_3 T_R^2}{m_S}  .
\end{equation}
After reheating, the field value evolves as
\begin{equation}
\label{eq:VSTquad}
v_{S}(T) = v_S(T_R) \left(\frac{T}{T_R}\right)^{ \scalebox{1.01}{$\frac{3}{2}$} }  =  \delta \left( \frac{\pi^2 g_*}{90}\right)^{ \scalebox{1.01}{$\frac{1}{8}$} } \frac{\alpha_3 T^{3/2} T_R^{1/2}}{m_S}
\end{equation}
and reaches $v_S$ at the temperature
\begin{equation}
T_S = T_R \left( \frac{v_S}{v_S(T_R)} \right)^{ \scalebox{1.01}{$\frac{2}{3}$} } \simeq 8 \times 10^7 \, \GeV  \left( \frac{k}{\delta} \right)^{ \scalebox{1.01}{$\frac{2}{3}$} } \left( \frac{10^{10} \, \GeV}{T_R} \right)^{ \scalebox{1.01}{$\frac{1}{3}$} } \left(\frac{m_S}{300 \, \GeV}\right)^{ \scalebox{1.01}{$\frac{2}{3}$} } \left( \frac{\rm TeV}{m_3} \right)^{ \scalebox{1.01}{$\frac{2}{3}$} } \left( \frac{m_{3/2}}{\GeV} \right)^{ \scalebox{1.01}{$\frac{2}{3}$} } .
\end{equation}

The sgoldstino eventually delivers all its energy to radiation by scattering with (decaying to) thermal particles at the destruction (decay) temperature $T_{\rm des}$ ($T_{\rm dec}$), which is discussed in Sec.~\ref{subsec:desS} (\ref{subsec:decayS}). 

\begin{figure}[tb]
 \begin{center}
  \includegraphics[width=0.8\linewidth]{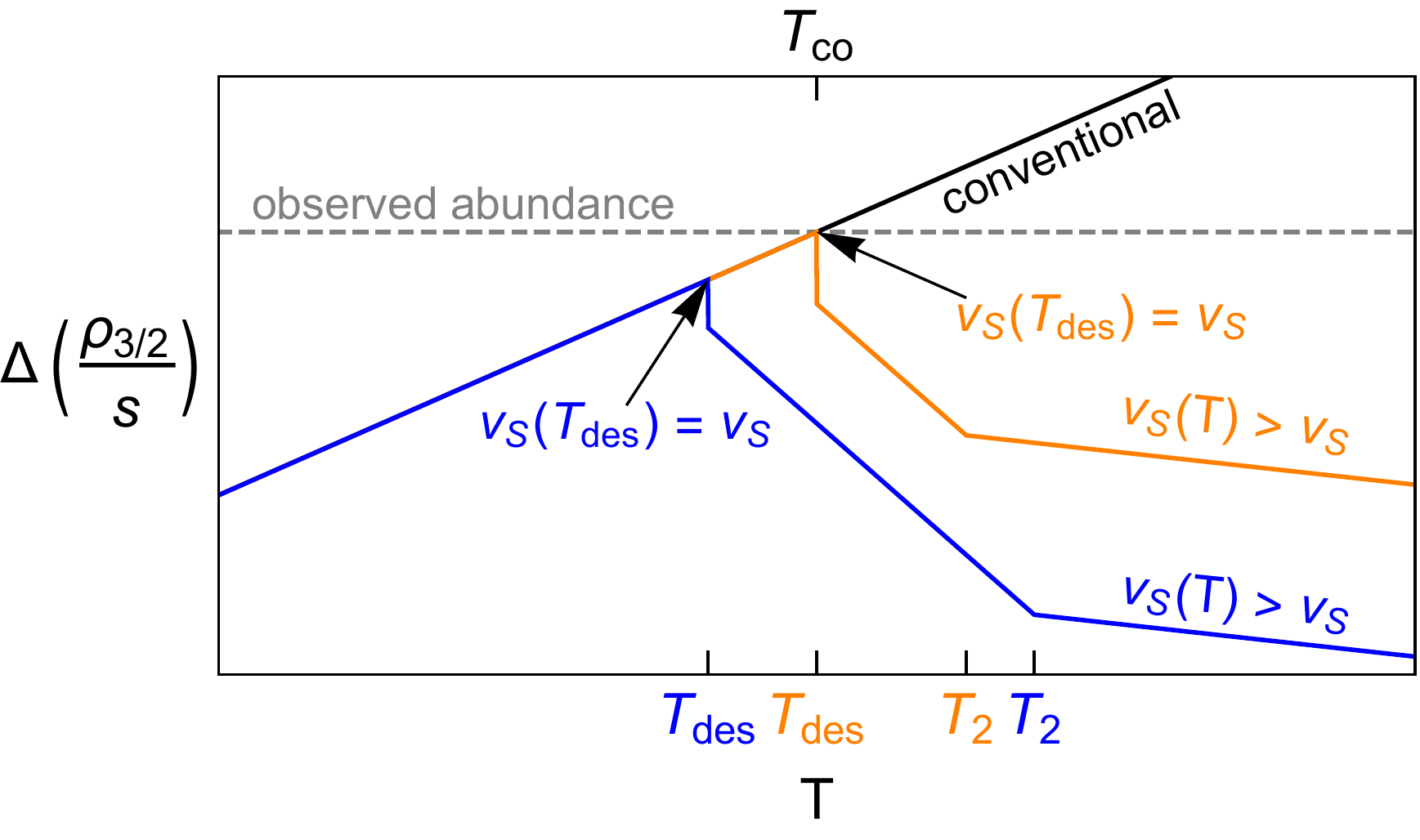}
 \end{center}
\caption{
This schematic diagram with log scale axes shows the suppression on gravitino production at high temperatures. The vertical axis is the gravitino abundance produced per Hubble time. The horizontal dashed line refers to the dark matter abundance observed today. The black and colored lines refer to the conventional and suppressed production respectively.}
\label{fig:schem}
\end{figure}

Fig.~\ref{fig:schem} summarizes the sgoldstino evolution and serves as the schematic picture of the suppression mechanism.
The vertical axis shows the gravitino abundance produced per Hubble time at a given temperature in the horizontal axis.
The conventional production is UV dominated and thus we need $T_R \le T_{\rm co}$ to avoid overproduction. In the case where $v_S(T)$ is decreasing faster than $T^{1/2}$, the suppressed production becomes IR dominated, as seen in high temperature behavior.
The kinks of the colored lines occur at $T_2$, as the temperature dependence of $v_S(T)$ changes from Eq.~(\ref{eq:VSTlog}) to Eq.~(\ref{eq:VSTquad}) at the transition from the thermal logarithmic potential to the quadratic one. The abrupt change of the lines at $T_{\rm des}$ orginates from the fact that the field value of the sgoldstino is suddenly set to the vev today when the condensate is destroyed by thermal scattering.
In the case of the orange line, the observed dark matter abundance is reproduced. On the other hand, the blue line represents underproduction.

\subsection{Destruction of Sgoldstinos by Thermal Scattering}
\label{subsec:desS}

The discussion in Sec.~\ref{sec:SgDyn} assumes that the sgoldstino condensate is intact throughout its evolution. However, due to its coupling with the messenger $Q$, the sgoldstino scatters with thermalized particles at the following rate given in Refs.~\cite{Mukaida:2012qn, Bodeker:2006ij}
\begin{equation}
\Gamma_{\rm scatt} = \left( \frac{T(Q)}{16\pi^2}  \right)^2 \frac{(12\pi)^2}{\ln(\alpha_3^{-1})}  \frac{\alpha_3^2 T^3}{v_S^2(T)} \equiv b \frac{\alpha_3^2 T^3}{v_S^2(T)} ,
\end{equation}
where $T(Q)$ is the index of $Q$'s representation of $SU(3)$ and we take $T(Q) = 1/2$. The condensate is destroyed whenever the scattering rate becomes larger than the Hubble rate. The temperature at which such destruction occurs is called $T_{\rm des}$.

\subsubsection{Sgoldstino Oscillations Driven by Thermal Effects}
\label{subsec:oscTherm}

We first explore the case where the sgoldstino begins to oscillate via the thermal logarithmic potential.
Overproduction of gravitinos excludes the possibility where the sgoldstino condensate is destroyed before the quadratic potential dominates, i.e. $T_{\rm des} > T_2$. This is because for such a case the field $S$ is trapped at the origin, making $Q$ massless and thermalized and greatly enhancing the gravitino production rate.%
\footnote{If the Yukawa coupling $y$ is sufficiently small, the produced gravitino abundance is not necessarily very large and thus the entropy production by the sgoldstino trapped at the origin may have sufficient dilution for the gravitino abundance~\cite{Fukushima:2012ra}. We however do not consider this scenario in this paper.}
Requiring $\Gamma_{\rm scatt}(T_2) < H(T_2)$ gives
\begin{equation}
T_R \lsim  10^{14} \, \GeV \, \delta^4 \left(\frac{m_S}{300 \, \GeV}\right)^{ \scalebox{1.01}{$\frac{1}{3}$} } .
\label{eq:TRT2}
\end{equation}
As the condensate is destroyed, the sgoldstino is driven to the local minimum of the potential. In order for $S=v_S$ to be the local minimum at the temperature $T_{\rm des}$, the thermal mass from the messenger should be small enough, \begin{equation}
y \, T_{\rm des}<m_S .
\label{eq:yTdes}
\end{equation}
This upper bound on $y$ should be consistent with the lower bound in Eq.~(\ref{eq:ymin}).

Below $T_2$, on the other hand, $\Gamma_{\rm scatt}(T)/H$ is IR dominated only before $T_S$. This implies either that $T_{\rm des} > T_S$ or that there is no destruction by scattering. In order to distinguish our mechanism from other solutions of the gravitino problem, we first explore the parameter space where the sgoldstino condensate does not produce entropy.
We take $v_S(T) = v_S (T/T_S)^{3/2}$ because $T_2 > T_{\rm des} > T_S$ and derive the destruction temperature
\begin{equation}
T_{\rm des} \simeq  6 \times 10^5 \, \GeV \, \delta^{-2} \left( \frac{T_R}{10^{12} \, \GeV} \right)^{ \scalebox{1.01}{$\frac{1}{2}$} } \left(\frac{m_S}{300 \, \GeV}\right)^{ \scalebox{1.01}{$\frac{1}{2}$} } .
\end{equation}
To be consistent with $T_{\rm des} > T_S$ so that the sgoldstino is successfully destroyed, one requires
\begin{equation}
k \lsim 10^{-4} \, \delta^{-1} \left( \frac{T_R}{10^{12} \, \GeV} \right)^{ \scalebox{1.01}{$\frac{1}{4}$} } \left(\frac{m_S}{300 \, \GeV}\right)^{ \scalebox{1.01}{$\frac{1}{4}$} } \left( \frac{m_3}{\rm TeV} \right)  \left( \frac{\GeV}{m_{3/2}} \right) .
\label{eq:kMin}
\end{equation}
In the case where $k < 1$, there should be another SUSY breaking field. If the scalar component of that SUSY breaking field is excited in the early Universe, its decay may also produce gravitinos. To avoid cosmological complications, we assume that this scalar component has a positive Hubble-induced mass and/or efficiently decays into hidden sector fields other than the gravitino.

According to Eq.~(\ref{eq:TRcons}), it is required that $T_{\rm des} \le T_{\rm co}$ to avoid overproduction of gravitinos. This  condition is satisfied when
\begin{equation}
T_R \lsim  10^{13} \, \GeV \, \delta^4 \left(\frac{300 \, \GeV}{m_S}\right) \left( \frac{m_{3/2}}{\GeV} \right)^2 \left( \frac{\rm TeV}{m_3} \right)^4.
\end{equation}
Furthermore, to identify the parameter space with no dilution, we need to ensure that the sgoldstino condensate is destroyed before its energy density dominates over radiation. We can estimate the temperature $T_M^{\rm (th)}$ at which the matter energy density dominates over that of radiation
\begin{equation}
\frac{\pi^2}{30} g_* (T_M^{\rm (th)})^4 = m_S^2 v_S^2(T_2) \left( \frac{T_M^{\rm (th)}}{T_2} \right)^3
\end{equation}
and the result reads
\begin{equation}
T_M^{\rm (th)} = \frac{30}{\pi^2} \frac{\alpha_3^2}{g_*} T_2 = \delta^4  \frac{30}{\pi^2} \frac{\alpha_3^2}{g_*} \frac{m_S M_{Pl}}{T_R} \simeq  10^5 \, \GeV \, \delta^4 \left( \frac{10^{12} \, \GeV}{T_R} \right) \left(\frac{m_S}{300 \, \GeV}\right) ,
\label{eq:TMth}
\end{equation}
where we use $\alpha_3^2 T_2^4 = m_S^2 v_S^2(T_2)$ (the definition of $T_2$). No entropy is produced when $T_{\rm des} > T_M^{\rm (th)}$, which is the case for
\begin{equation}
T_R \gsim 10^{11} \, \GeV \, \delta^4 \left(\frac{m_S}{300 \, \GeV}\right)^{ \scalebox{1.01}{$\frac{1}{3}$} }  .
\end{equation}

If the scattering is inefficient, the sgoldstino dominates the energy density of the Universe.
After the sgoldstino dominates, destruction occurs via scattering with the thermal bath created from the condensation of the sgoldstino. The destruction temperature is derived in App.~\ref{app:TdesMD} and reads
\begin{equation}
T_{\rm des}   \simeq 3 \times 10^5 \, \GeV \left( \frac{m_S}{300 \, \GeV} \right)^{ \scalebox{1.01}{$\frac{2}{3}$} }.
\label{eq:Tdes_text}
\end{equation}
The dilution factor $D$ via entropy production is given by
\begin{equation}
D = \frac{T_M^{\rm (th)}}{T_{\rm des}} \simeq 3 \, \delta^4\left( \frac{m_S}{300 \, \GeV} \right)^{ \scalebox{1.01}{$\frac{1}{3}$} } \left( \frac{10^{11}\, \GeV}{T_R} \right).
\end{equation}
The condition from Eq.~(\ref{eq:TRcons}) is $T_{\rm des} /D \le T_{\rm co}$ so the upper bound on $T_R$ is relaxed to
\begin{equation}
T_R \lsim 10^{12} \, \GeV \, \delta^4 \left( \frac{m_{3/2}}{\GeV} \right) \left( \frac{300 \, \GeV}{m_S} \right)^{ \scalebox{1.01}{$\frac{1}{3}$} } \left( \frac{\rm TeV}{m_3} \right)^{2}. 
\label{eq:TdesDTco1}
\end{equation}

\subsubsection{Sgoldstino Oscillations Driven by Vacuum Potential}
\label{subsec:oscVac}

We next explore the case where the sgoldstino begins to oscillate via the vacuum mass term.
In the case where the condensate is a subdominant component, the destruction temperature is given by
\begin{equation}
T_{\rm des} = \frac{3^{3/4} 10^{3/8} \sqrt{b} }{\pi ^{3/4} \delta g_*^{3/8} } \frac{m_S \sqrt{M_{\text{Pl}}}}{\sqrt{T_R}} 
\simeq 3 \times 10^5 \, \GeV  \delta^{-1} \left( \frac{m_S}{300 \, \GeV} \right) \left( \frac{10^9 \, \GeV}{T_R} \right)^{ \scalebox{1.01}{$\frac{1}{2}$} } .
\end{equation}
One needs to require $T_{\rm des} > T_S$ to ensure successful destruction, which limits 
\begin{equation}
k \lsim 10^{-4} \, \delta^{-1/2} \left( \frac{m_S}{300 \, \GeV} \right)^{ \scalebox{1.01}{$\frac{1}{2}$} } \left( \frac{m_3}{\rm TeV} \right) \left( \frac{\GeV}{m_{3/2}} \right) \left( \frac{10^9 \, \GeV}{T_R} \right)^{ \scalebox{1.01}{$\frac{1}{4}$} }.
\label{eq:kMin2}
\end{equation}
When the scattering rate is inefficient, the Universe enters the matter-dominated era at the temperature 
\begin{equation}
T_M^{\rm (vac)} = \frac{v_{S0}^2 T_R}{3 M_{Pl}^2} = \left( \frac{90}{\pi^2g_*} \right)^{ \scalebox{1.01}{$\frac{3}{4}$} } \frac{\delta^2 \alpha_3^2 T_R}{3} .
\label{eq:TMvac}
\end{equation}
With the destruction temperature given in Eq.~(\ref{eq:Tdes}), the dilution factor can be computed
\begin{equation}
D = \frac{T_M^{\rm (vac)}}{T_{\rm des}} = \frac{ 10^{1/4} \alpha _3^{4/3} \delta^2 }{ 3^{1/6} \sqrt{\pi} g_*^{1/4}  b^{1/3} } \frac{T_R}{\left( M_{Pl} m_S^2 \right)^{1/3}} \simeq 9 \, \delta^2   \left( \frac{T_R}{10^{10} \, \GeV} \right) \left( \frac{300 \, \GeV}{m_S} \right)^{ \scalebox{1.01}{$\frac{2}{3}$} } .
\end{equation}
Therefore, the sgoldstino does not produce entropy when 
\begin{equation}
T_R \lsim 10^9 \, \GeV \delta^{-2}  \left( \frac{m_S}{300 \, \GeV} \right)^{ \scalebox{1.01}{$\frac{2}{3}$} } .
\end{equation} 
The condition from Eq.~(\ref{eq:TRcons}) is $T_{\rm des} /D \le T_{\rm co}$ and places the following upper bound on $T_R$
\begin{equation}
T_R \lsim 2 \times 10^{10} \, \GeV \, \left(\frac{0.6}{\delta}\right)^2 \left( \frac{200 \rm \, MeV}{m_{3/2}} \right) \left( \frac{m_S}{\rm TeV} \right)^{ \scalebox{1.01}{$\frac{4}{3}$} } \left( \frac{m_3}{2 \, \rm TeV} \right)^{2}. 
\label{eq:TdesDTco2}
\end{equation}

Let us discuss the compatibility with thermal leptogenesis. The maximal baryon asymmetry $Y_{B, \rm max}$ that can be obtained from thermal leptogenesis in the units of that observed today $Y_{B, \rm obs}$ is given in Refs.~\cite{Giudice:2003jh, Buchmuller:2004nz}
\begin{equation}
Y_{B, \rm max} \simeq \frac{T_R}{10^9 \, \GeV} Y_{B, \rm obs}.
\end{equation}
This baryon asymmetry may be subject to dilution from subsequent entropy production, which leads to a more stringent lower bound on $T_R$,
\begin{equation}
\frac{T_R}{D(T_R)} \gsim  10^9 \, \GeV.
\label{eq:TRleptogen}
\end{equation}

\begin{figure}[tb]
 \begin{center}
  \includegraphics[width=0.49\linewidth]{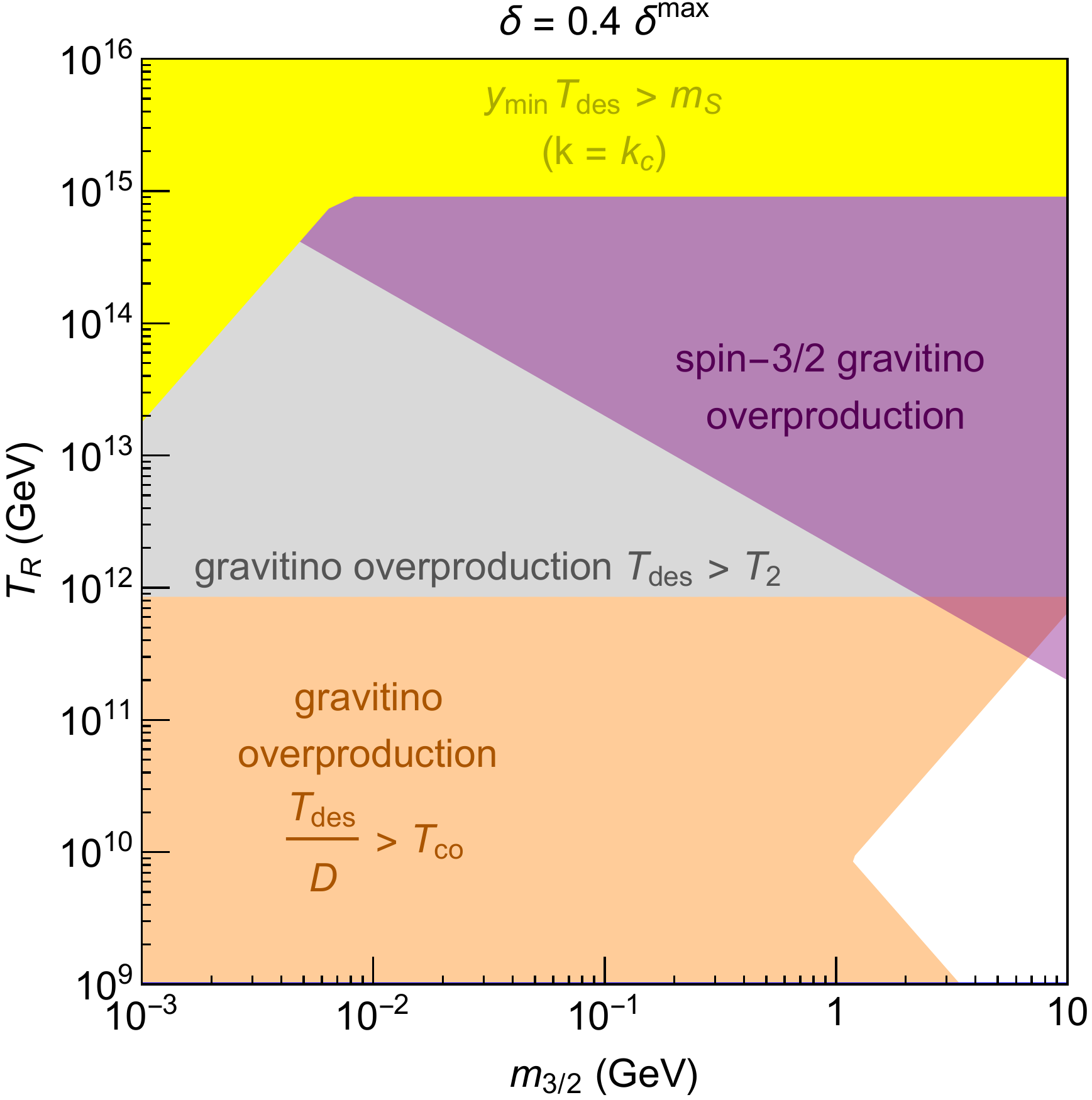}
  \includegraphics[width=0.49\linewidth]{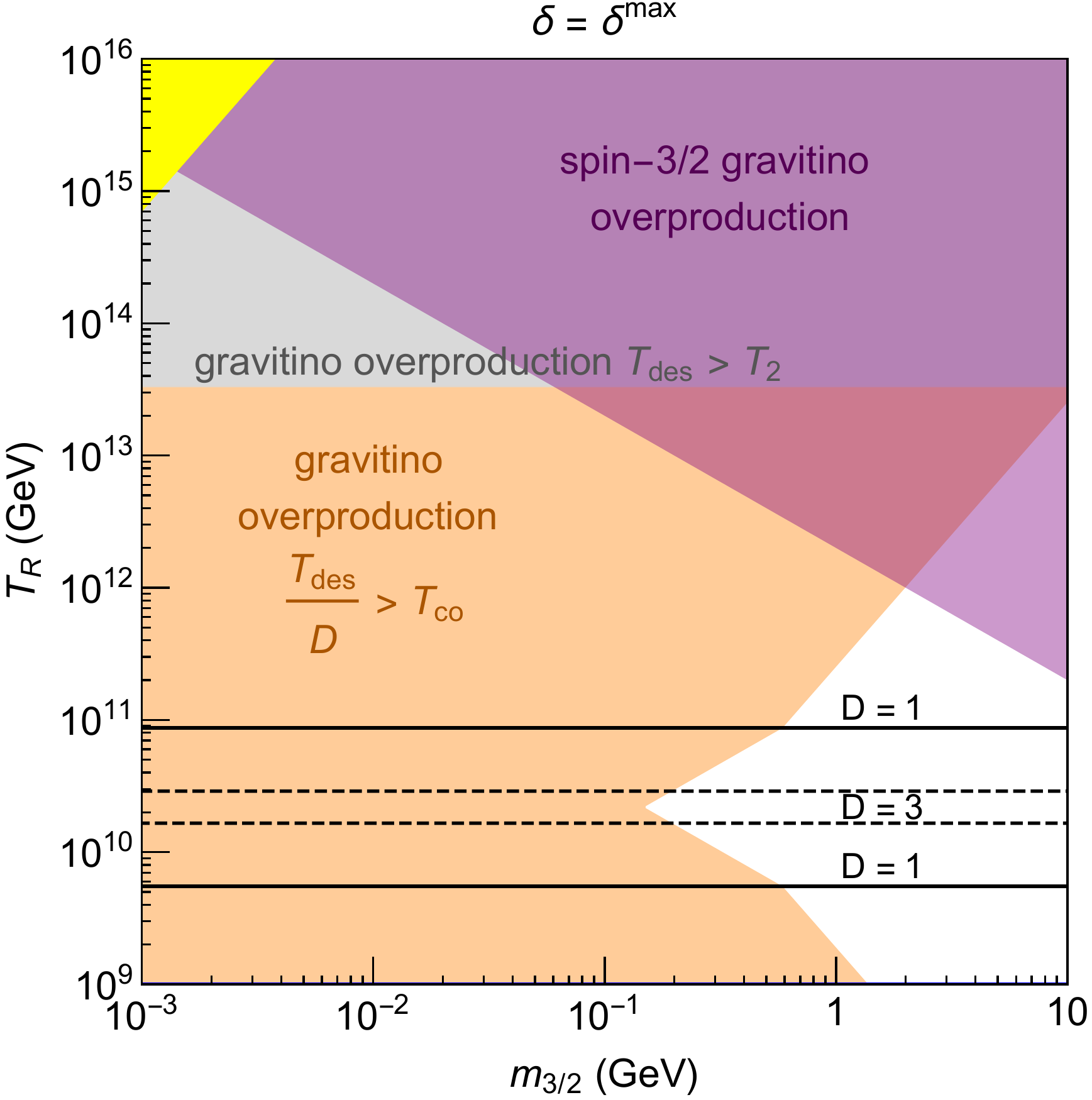}
 \end{center}
\caption{
The various constraints on the reheat temperature and the gravitino mass are shown as the shaded regions. These panels apply to the theories with indirect gauge mediation where the $F$ term of $S$ provides a small amount of supersymmetry breaking, as required in Eq.~(\ref{eq:kMin}). The energy of the sgoldstino condensate is transferred to radiation by thermal scattering. Leptogenesis can provide the observed baryon asymmetry in the entire allowed (unshaded) region. We take $m_3 = 2$ TeV and $m_S = 1$ TeV. }
\label{fig:MasterScatt}
\end{figure}

There is a further constraint from the production of the fermion component of $S$, $\psi_S$ for the following reason. Since we are currently concerned with the case where the sgoldstino condensate is destroyed by thermal scattering, a small $k$ given in Eq.~(\ref{eq:kMin}) is assumed.
For a small $k$, the production of $\psi_S$ is enhanced by $1/k^2$ compared to that of the gravitino.
To avoid the gravitino overproduction from the decay of $\psi_S$, we require that the mass of $\psi_S$ is larger than that of the lightest observable supersymmetric partner (LOSP) so that $\psi_S$ can decay into the LOSP.%
\footnote{This requires a direct coupling between the SUSY breaking sector containing $S$ and other SUSY breaking sector. Otherwise, $\psi_S$ is a pseudo-goldstino and obtains a mass only as large as the gravitino.}
We find that the LOSP from the decay of $\psi_S$ immediately annihilate to the SM particles with a negligible amount decaying to the gravitino. 
The mass of the sgoldstino would not be much smaller than that of $\psi_S$, and thus we fix $m_S = 1$ TeV. 

We summarize the above discussions in Fig.~\ref{fig:MasterScatt}. The light gray region is excluded by Eq.~(\ref{eq:TRT2}) because the sgoldstino is destroyed by thermal scattering when the potential is still governed by the thermal logarithmic potential, whose minimum is at the origin of $S$. This results in a vanishing messenger scale and the gravitino is overproduced by the scattering of thermalized messengers. The orange region is excluded by Eqs.~(\ref{eq:TdesDTco1}) and (\ref{eq:TdesDTco2}) because the sgoldstino destruction occurs too early and the mechanism fails to suppress the messenger scale.
The dilution factor $D$ is labeled by the black contours in the right panel, whereas the left panel does not have dilution for the chosen value of $\delta$. The maximal value $\delta_{\rm max} \simeq 0.67$ is inferred from Eqs.~(\ref{eq:deltaDef}) and (\ref{eq:condition_thermal}). Above (below) the sharp kink of the orange boundary, the onset of the sgoldstino oscillation is driven by the thermal logarithmic (vacuum) potential, discussed in Secs.~\ref{subsec:oscTherm} and \ref{subsec:oscVac} respectively. The yellow region is excluded by Eq.~(\ref{eq:yTdes}) because, at the time of sgoldstino destruction, the thermal mass dominates and the destruction will set the field value to the origin, resulting in the gravitino overproduction. Here $k_c$ refers to the critical value in Eq.~(\ref{eq:kMin}) for successful destruction. The purple region is excluded by Eq.~(\ref{eq:TRspin32}) for overproduction of spin-3/2 gravitinos through supergravity interactions.
We find that in the allowed parameter region the thermal leptogenesis can create an enough amount of the lepton asymmetry.

\subsection{Destruction of Sgoldstinos by Decay}
\label{subsec:decayS}

It is pointed out in Sec.~\ref{subsec:desS} that $k$ has to be smaller than the critical value $k_c$ given in Eqs.~(\ref{eq:kMin}), (\ref{eq:kMin2}) or (\ref{eq:kMin3}) in order for the sgoldstino condensate to be destroyed by thermal scattering. In this section, we assume a sufficiently large $k$, meaning that thermal scattering is never effective enough and instead the sgoldstino condensate eventually decays to particles in the thermal bath. 

The real and imaginary parts of $S$ may have different decay modes \cite{Hamaguchi:2009hy}. 
In the phase convention where $v_S$ is real,
both the real and imaginary components of $S$ can decay to a pair of gluons at a rate
\begin{equation}
\Gamma_{\rm dec}^{gg} \simeq \left( \frac{\alpha_3}{4\pi} \right)^2 \frac{m_S^3}{8\pi v_S^2}.
\end{equation}
The real component of $S$ can also decay to Higgs/electroweak (EW) gauge bosons if kinematically allowed at a rate
\begin{equation}
\Gamma_{\rm dec}^{\rm h,W,Z} \simeq  \frac{1}{8\pi}  \frac{m_H^4}{m_S v_S^2} .
\end{equation}
Assuming $m_S\sim$ TeV, this decay mode is more efficient than the one into gluons.
The relative abundance of the real and imaginary parts depends on the phase of the initial field value $v_{S0}$. As the decay to Higgs is more efficient, the real component will decay before the imaginary one. As a result, the final decay temperature is mainly governed by the decay to gluons if the initial relative abundance is comparable or dominated by the imaginary component.

To find the temperature $T_{\rm dec}$ when the sgoldstino decays to gluons, one equates the decay rate with the Hubble rate and obtains
\begin{equation}
T_{\rm dec}^{gg} \simeq \sqrt{\Gamma_{\rm dec}^{gg} M_{Pl}} \simeq 
4 \, {\rm MeV} \, k^{-1} \left( \frac{m_S}{300 \, \GeV} \right)^{ \scalebox{1.01}{$\frac{3}{2}$} } \left( \frac{\rm 100 \, MeV}{m_{3/2}} \right) \left( \frac{m_3}{\rm TeV} \right) .
\end{equation}
The decay temperature then allows us to compute the dilution factor using Eq.~(\ref{eq:TMth})
\begin{equation}
D = \frac{T_M^{\rm (th)}}{T_{\rm dec}^{gg}} \simeq 
8 \times 10^6 \, k \, \delta^4 \left( \frac{10^{12} \, \GeV}{T_R} \right) \left( \frac{300 \, \GeV}{m_S} \right)^{ \scalebox{1.01}{$\frac{1}{2}$} } \left( \frac{m_{3/2}}{\rm 100 \, MeV} \right) \left( \frac{\rm TeV}{m_3} \right) .
\label{eq:DdecGluons}
\end{equation}
Since the decay occurs well after the field value of $S$ settles to the minimum today $v_S$, the gravitino production peaks at $T_S$. With dilution, the constraint from the gravitino abundance using Eq.~(\ref{eq:TRcons}) becomes $T_S /D \le T_{\rm co}$ and gives
\begin{equation}
T_R \lsim 
5 \times 10^{15} \, \GeV \, k^{1/4} \, \delta^4 \left( \frac{300 \, \GeV}{m_S} \right)^{ \scalebox{1.01}{$\frac{5}{8}$} } \left( \frac{m_{3/2}}{\rm 100 \, MeV} \right)  \left( \frac{\rm TeV}{m_3} \right)^{ \scalebox{1.01}{$\frac{7}{4}$} } .
\label{eq:TSDTco1}
\end{equation}

With accidental suppression of the imaginary part of the initial field value or the presence of a CP violating mixing between the real and imaginary components of $S$,
the decay temperature $T_{\rm dec}$ is now determined by the larger of the rate into gluons and that into EW bosons. For the decays to $H$, $W^\pm$, and $Z$, we obtain
\begin{equation}
T_{\rm dec}^{\rm h,W,Z} \simeq \sqrt{\Gamma_{\rm dec}^{\rm h,W,Z} M_{Pl}} \simeq 
5 \, \GeV \, k^{-1} \left( \frac{m_H}{\rm TeV} \right)^2 \left( \frac{300 \, \GeV}{m_S} \right)^{ \scalebox{1.01}{$\frac{1}{2}$}} \left( \frac{\rm 100 \, MeV}{m_{3/2}} \right) \left( \frac{m_3}{\rm TeV} \right).
\end{equation}
 The dilution factor becomes
\begin{equation}
D = \frac{T_M^{\rm (th)}}{T_{\rm dec}^{\rm h,W,Z}} \simeq 
6 \times 10^3 \, k \, \delta^4 \left( \frac{10^{12} \, \GeV}{T_R} \right) \left( \frac{\rm TeV}{m_H} \right)^2 \left( \frac{m_S}{300 \, \GeV} \right)^{ \scalebox{1.01}{$\frac{3}{2}$}} \left( \frac{m_{3/2}}{\rm 100 \, MeV} \right) \left( \frac{\rm TeV}{m_3} \right) .
\label{eq:DdecHiggs}
\end{equation}
Finally, the constraint of the gravitino abundance from Eq.~(\ref{eq:TRcons}) requires $T_S /D \le T_{\rm co}$, giving
\begin{equation}
T_R \lsim 
3 \times 10^{13} \, \GeV \, k^{1/4} \, \delta^4 \left( \frac{\rm TeV}{m_H} \right)^{ \scalebox{1.01}{$\frac{3}{2}$}} \left( \frac{m_S}{300 \, \GeV} \right)^{ \scalebox{1.01}{$\frac{7}{8}$}} \left( \frac{m_{3/2}}{\rm 100 \, MeV} \right)  \left( \frac{\rm TeV}{m_3} \right)^{ \scalebox{1.01}{$\frac{7}{4}$}}.
\label{eq:TSDTco2}
\end{equation}
In the case where the condition in Eq.~(\ref{eq:condition_thermal}) is violated, the sgoldstino starts oscillating via the vacuum potential $V_{\rm vac}(S)$ rather than the thermal potential $V_{\rm th}(S)$. The decay temperatures $T_{\rm dec}$ calculated above do not change but the matter-domination temperature should become $T_M^{\rm (vac)}$ in Eq.~(\ref{eq:TMvac}). The dilution factors are modified as 
\begin{align}
D = & \, \frac{T_M^{\rm (vac)}}{T_{\rm dec}^{\rm gg}} \simeq 8 \times 10^8 \, k \, \delta^2 \left( \frac{T_R}{10^{10} \, \GeV} \right) \left( \frac{300 \, \GeV}{m_S} \right)^{ \scalebox{1.01}{$\frac{3}{2}$}} \left( \frac{m_{3/2}}{\rm 100 \, MeV} \right) \left( \frac{\rm TeV}{m_3} \right) , \\
D = & \, \frac{T_M^{\rm (vac)}}{T_{\rm dec}^{\rm h,W,Z}} \simeq 6 \times 10^5 \, k \, \delta^2 \left( \frac{T_R}{10^{10} \, \GeV} \right) \left( \frac{m_S}{300 \, \GeV} \right)^{ \scalebox{1.01}{$\frac{1}{2}$}} \left( \frac{m_{3/2}}{\rm 100 \, MeV} \right) \left( \frac{\rm TeV}{m_3} \right) \left( \frac{\rm TeV}{m_H} \right)^2 ,
\end{align}
for the decays to into gluons and into Higgs and EW bosons, respectively.

In addition, the sgoldstino can decay to a pair of gravitinos at a rate given by Ref.~\cite{Ibe:2006rc}
\begin{equation}
\Gamma_{\rm dec}^{3/2} = \frac{k^2 m_S^5}{96 \pi^2 m_{3/2}^2 M_{Pl}^2},
\end{equation}
where $k$ accounts for the mixing of the gravitino with $\psi_S$. Based on Eqs.~(\ref{eq:DdecGluons}) and (\ref{eq:DdecHiggs}), the sgoldstino dominates the energy density in the parameter space of interest. This implies that the sgoldstino decay can give a sizable contribution to the gravitino despite the small branching ratio $B_{3/2} \equiv \Gamma_{\rm dec}^{3/2}  / \Gamma_{\rm tot}$. For the decays to gluinos and to $H$, $W^\pm$, and $Z$ respectively, the non-thermal gravitino abundance is estimated as
\begin{align}
\label{eq:NonThGravi}
\frac{\rho_{3/2}}{s} = & \  2 m_{3/2} \frac{\rho_S B_{3/2}}{m_S s} = \frac{k^2 m_S^4}{48 \pi m_{3/2} M_{Pl} T_{\rm dec}}  \nonumber \\
 \simeq & 
\begin{cases}
0.6 \, {\rm eV} \left( \frac{k}{0.1} \right)^3 \left( \frac{m_S}{300 \, \GeV} \right)^{ \scalebox{1.01}{$\frac{5}{2}$} } \left( \frac{\rm TeV}{m_3} \right) \\
0.04 \, {\rm eV} \, k^3 \left( \frac{\rm TeV}{m_H} \right)^2 \left( \frac{m_S}{300 \, \GeV} \right)^{ \scalebox{1.01}{$\frac{9}{2}$} }  \left( \frac{\rm TeV}{m_3} \right) .
\end{cases}
\end{align} 

\begin{figure}
 \begin{center}
  \includegraphics[width=0.49\linewidth]{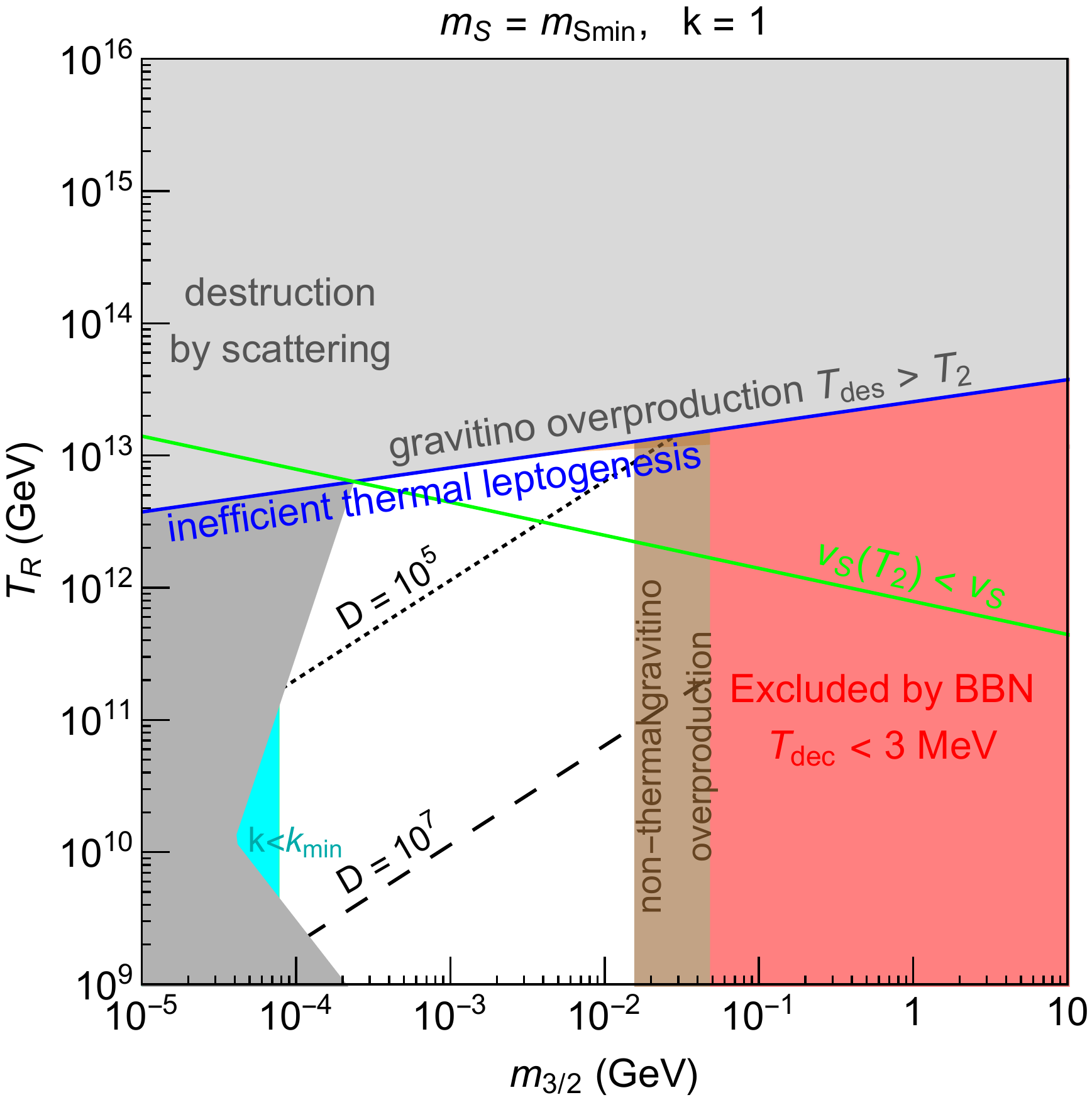}   \includegraphics[width=0.49\linewidth]{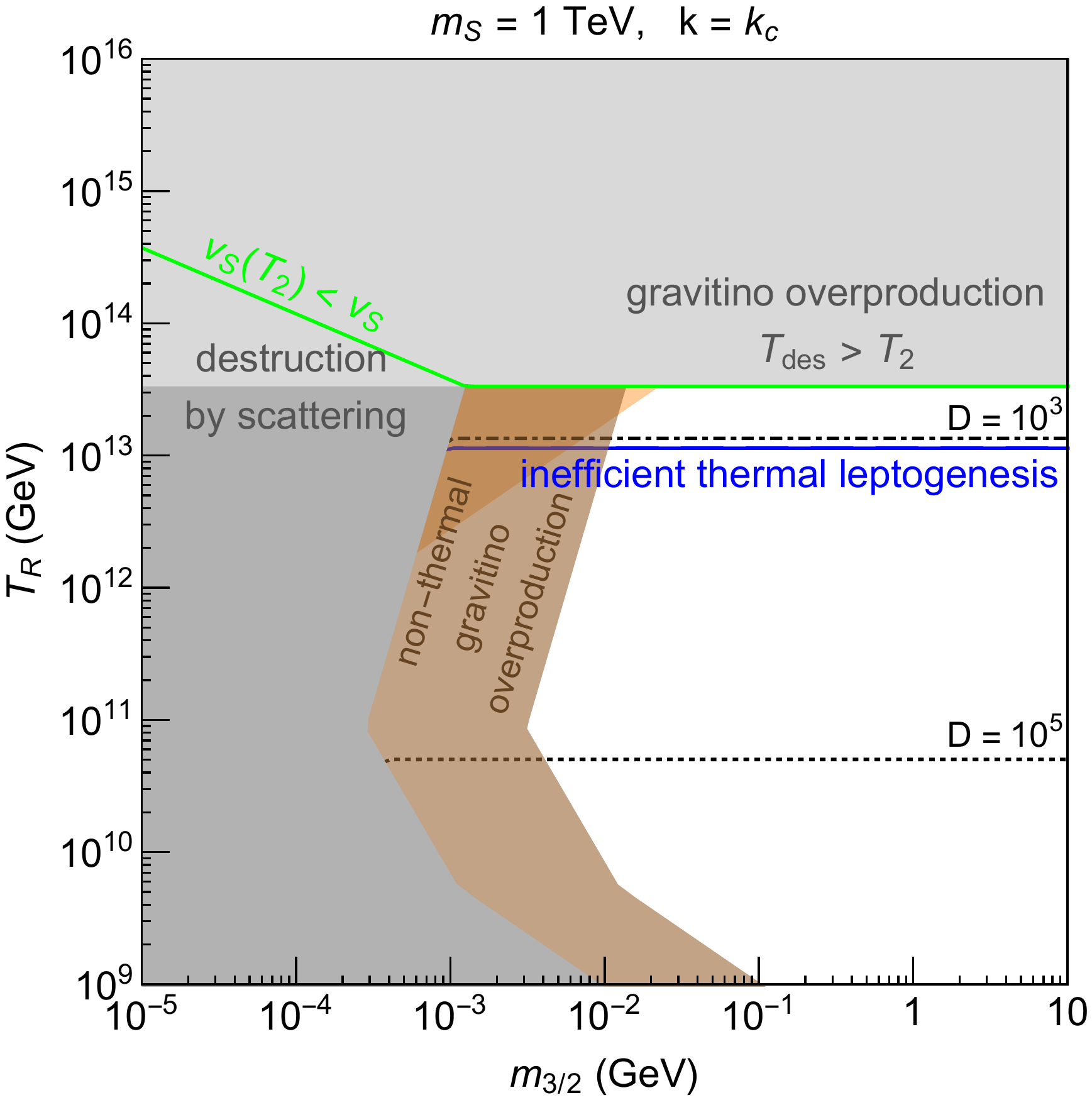}
  \includegraphics[width=0.49\linewidth]{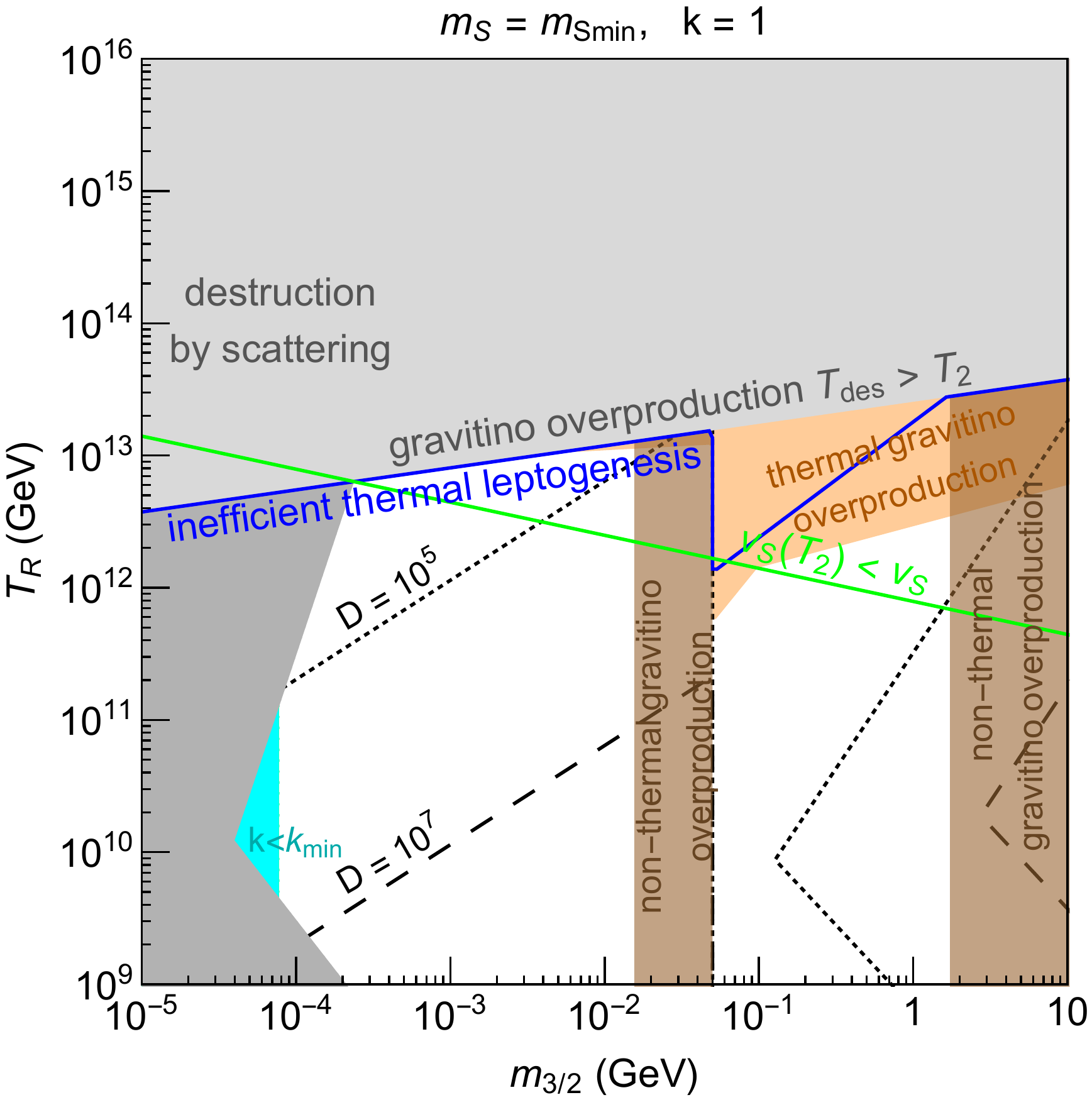}   \includegraphics[width=0.49\linewidth]{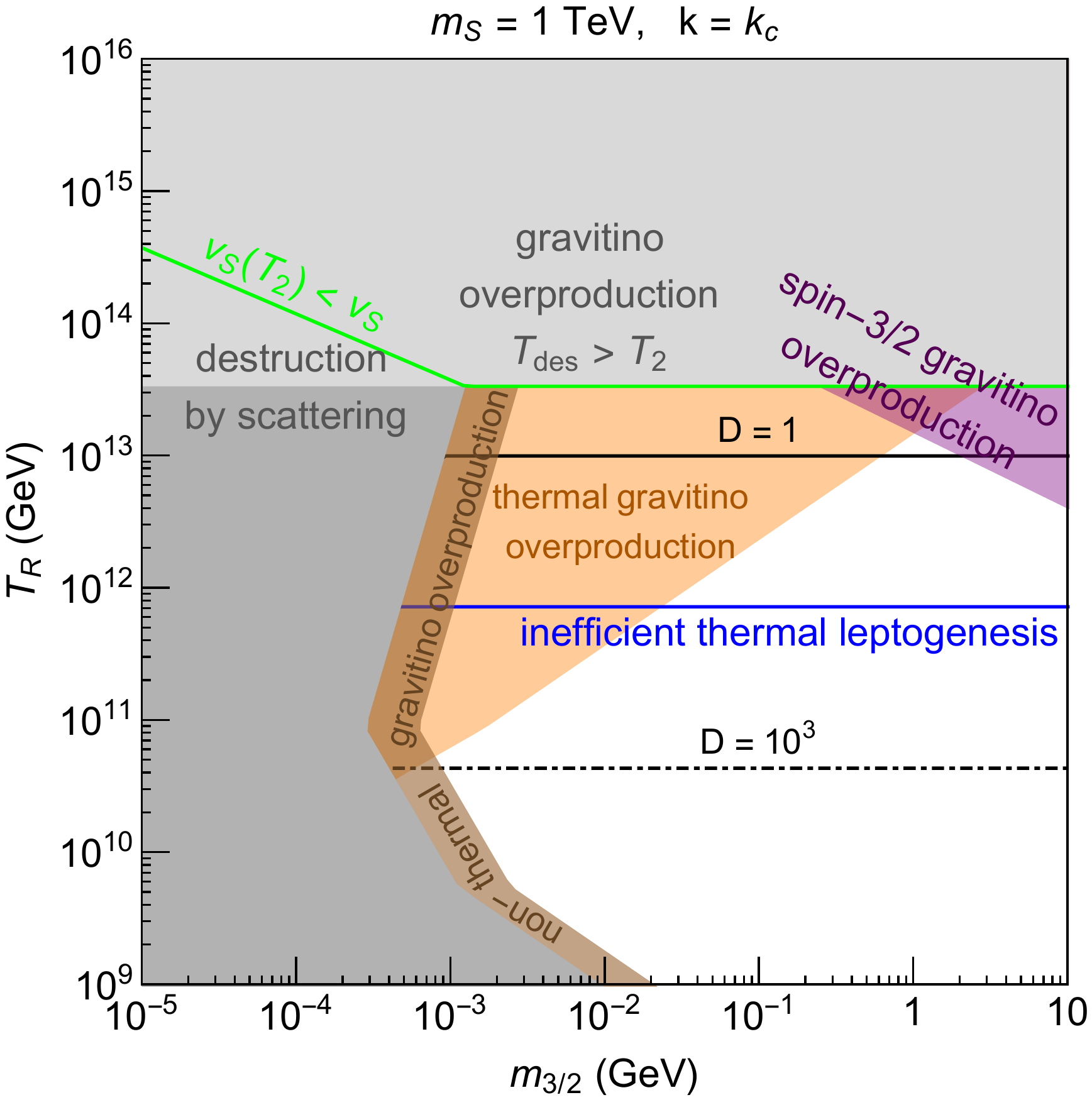}
 \end{center}
\caption{
The various constraints on the reheat temperature and the gravitino mass are shown as the shaded regions. The left (right) panels apply to the theories with (direct) indirect gauge mediation where the $F$ term of $S$ provides a full (small) amount of supersymmetry breaking. The sgoldstino condensate eventually decays to gluons (and Higgs/EW bosons) in the upper (lower) two panels. Leptogenesis can provide the observed baryon asymmetry above the blue contours. We take $m_3 = 2$ TeV. }
\label{fig:MasterDecay}
\end{figure}

The various constraints discussed in this section are shown in Fig.~\ref{fig:MasterDecay}. In the left panels, we take $m_S = m_{S \rm min}$, which refers to the theoretical minimum given in Eq.~(\ref{eq:ymin}). $k_{\rm min}$ is the theoretical lower bound in Eq.~(\ref{eq:kmin}), which excludes the cyan region. In the right panels, we take $k = k_c$, where $k_c$ stands for the critical value of $k$ in Eqs.~(\ref{eq:kMin}), (\ref{eq:kMin2}), and (\ref{eq:kMin3}) applicable for different ranges of $T_R$. In the light (dark) gray region, the sgolstino is necessarily destroyed by scattering at high temperatures because $\Gamma_{\rm scatt}(T_2) > H(T_2)$ ($k_c > 1$), whose result is previously shown in Fig.~\ref{fig:MasterScatt}. We use the maximal value $\delta_{\rm max} \simeq 0.67$ inferred from Eqs.~(\ref{eq:deltaDef}) and (\ref{eq:condition_thermal}). The dilution factors in Eqs.~(\ref{eq:DdecGluons}) and (\ref{eq:DdecHiggs}) are marked with the black contours. The green contours separate different cosmological evolutions, where $v_S(T)$ in Eq.~(\ref{eq:vST2vS}) does (not) drop below $v_S$ above (below) the contours. The orange region is excluded by Eqs.~(\ref{eq:TSDTco1}) and (\ref{eq:TSDTco2}) because the sgoldstino field values drops to the today's value $v_S$ too quickly such that the mechanism fails to suppress the gravitino production until the conventional constraint temperature $T_{\rm co}$. The brown region is excluded by Eq.~(\ref{eq:NonThGravi}) because the gravitino produced from the sgoldstino decay overcloses the Universe. The purple region is excluded by Eq.~(\ref{eq:TRspin32}) due to overproduction of spin-3/2 gravitinos via supergravity interactions. The red regions are excluded as the decay of sgoldstinos occurs after and thus spoils Big Bang Nucleosynthesis (BBN) \cite{Kawasaki:2000en}. Regions below the blue contours are incompatible with thermal leptogenesis because the baryon asymmetry in Eq.~(\ref{eq:TRleptogen}) is depleted by too large of a dilution factor. The blue contours do not extend into the light gray regions, where the dilution factor is unity. For a smaller $\delta$, the orange region as well as the blue line shift downward. The lower bound on $T_R$ from thermal leptogenesis is then relaxed, until the orange region catches up with the blue line.

\section{Conclusions}
We have investigated the possibility that the sgoldstino has a large field value in the early Universe. This suppresses the early production of the gravitino and is expected to relax the upper bound on the reheat temperature after inflation. As a proof of principle, we analyze a specific case where the supersymmetry breaking field $S$ and the messenger fields couple minimally via Eq.~(\ref{eq:WSQQ}) and the mass term governs the zero-temperature potential of the sgoldstino.
The constraints on the gravitino mass and the reheat temperature are summarized in Figs.~\ref{fig:MasterScatt}-\ref{fig:MasterDecay}.
When the field $S$ provides sufficiently subdominant supersymmetry breaking, the sgoldstino condensate is destroyed by thermal scattering without producing (much) entropy. The reheat temperature may be as large as $10^{12}$~{\rm GeV}, and thermal leptogenesis is viable as long as the reheat temperature is larger than $10^9$ GeV.
On the contrary, if thermal scattering is inefficient, the sgoldstino condensate decays late with entropy production. The gravitino problem is then solved both by the suppression of the gravitino production and by dilution from entropy production. For a given reheat temperature, the dilution factor required to obtain a small enough gravitino abundance is smaller in our mechanism than the conventional scenario with dilution but not suppression. As a result, the reheat temperature can be as high as $3\times 10^{13}$ GeV. When the sgoldstino field breaks supersymmetry subdominantly and later decays, thermal leptogenesis is possible with a reheat temperature $T_R \gsim 10^{12\mathchar`-13}$ GeV. Hence, there exist regions in the parameter space where thermal leptogenesis is viable and the gravitino problem is absent or much milder than previously claimed.

\section*{Acknowledgement}
The authors thank Yiannis Dalianis, Lawrence Hall, and Aaron Pierce for fruitful discussions.
This work was supported in part by the Director, Office of Science, Office of High Energy and Nuclear Physics, of the US Department of Energy under Contract DE-AC02-05CH11231 and by the National Science Foundation under grants PHY-1316783 and PHY-1521446. The work of R.C. was in part supported by the U.S. Department of Energy, Office of Science, Office of Workforce Development for Teachers and Scientists, Office of Science Graduate Student Research (SCGSR) program. The SCGSR program is administered by the Oak Ridge Institute for Science and Education for the DOE under contract number DE-SC0014664. The work of K.H. was in part performed at the Aspen Center for Physics, which is supported by National Science Foundation grant PHY-1607611.

\appendix

\section{Destruction Temperature after Matter-Dominated Era}
\label{app:TdesMD}

In this section, we derive the destruction temperature in the case where the sgoldstino condensate dominates the energy of the Universe.
First of all, we need to require again that $v_S(T)$ is still decreasing before the condensate is destroyed, i.e. $T_S < T_{\rm des}$. Otherwise, the scattering rate is insufficient to destroy the sgoldstino condensate. Under this assumption, we first study the temperature dependence of the Hubble rate during the non-adiabatic era when the dominant source of the thermal bath is the scattering products of the sgoldstino as opposed to existing radiation.
We define the temperature at the beginning of the non-adiabatic era as $T_{\rm NA}$. By conservation of energy transferred from the sgoldstino condensate to radiation, we write
\begin{align}
\frac{\rho_S \Gamma_{\rm scatt} }{H} = & \ 3 M_{Pl}^2 H \Gamma_{\rm scatt} = \frac{\pi^2}{30} g_* T^4 , \\
\Gamma_{\rm scatt} = & \ b \frac{\alpha_3^2  T^3}{v_S^2(T)} = b \frac{\alpha_3^2 m_S^2 T^3}{3 H^2 M_{Pl}^2} ,
\end{align}
where we repeatedly use the fact that the total energy density is dominated by the sgoldstino vacuum potential given in Eq.~(\ref{eq:VvacS}), $H = \sqrt{\rho_S/3} / M_{Pl} = m_S v_S(T) / \sqrt{3}M_{Pl}$. The Hubble rate during this non-adiabatic phase is then given by
\begin{equation}
H(T) = \frac{30 \, b \, \alpha_3^2 }{\pi^2 g_* } \frac{m_S^2}{T}.
\end{equation}
This demonstrates that the Hubble rate is inversely proportional to the temperature and that the temperature during the non-adiabatic phase is increasing over time. As  $ \rho_S \propto H^2(T) \propto T^{-2}$, one can compare this new scaling with the usual temperature dependence $\rho_S \propto T^3$ during a radiation-dominated epoch and argue that the dilution factor is $D = T_M / T_{\rm des} = ( T_{\rm des} / T_{\rm NA})^5 $. We consider $T_{\rm des}$ as the temperature at which $H(T_{\rm des})$ is determined by the radiation energy density $\rho_R(T_{\rm des}) \propto T_{\rm des}^4$, which leads to
\begin{equation}
T_{\rm des} = \frac{ 3^{2/3} \sqrt{10} \alpha _3^{2/3} b^{1/3} }{\pi  \sqrt{g_*}} \left( M_{Pl} m_S^2 \right)^{ \scalebox{1.01}{$\frac{1}{3}$} } \simeq 3 \times 10^5 \, \GeV \left( \frac{m_S}{300 \, \GeV} \right)^{ \scalebox{1.01}{$\frac{2}{3}$} }.
\label{eq:Tdes}
\end{equation}
In fact, at the destruction temperature, the energy densities of the sgoldstino and radiation are comparable within a factor of a few, which allows us to compute the field value at $T_{\rm des}$
\begin{equation}
v_S(T_{\rm des}) = \sqrt{\frac{\pi^2 g_*}{30}} \frac{T_{\rm des}^2}{m_S} = \frac{3^{5/6} \sqrt{10} \alpha _3^{4/3} b^{2/3}}{\pi  \sqrt{g_*}} \left( m_S M_{Pl}^2 \right)^{ \scalebox{1.01}{$\frac{1}{3}$} }.
\end{equation}
The earlier assumption, $T_S < T_{\rm des}$, is equivalent to $v_S(T_{\rm des}) > v_S$, which places an upper bound on $k$
\begin{equation}
k \leq \frac{4 \times 3^{2/3} \sqrt{10} \alpha _3^{1/3} b^{2/3} }{\sqrt{g_*} } \frac{m_3}{m_{3/2}} \left( \frac{m_S}{M_{Pl}}\right)^{ \scalebox{1.01}{$\frac{1}{3}$} }
\simeq 10^{-4} \left( \frac{m_S}{300 \, \GeV} \right)^{ \scalebox{1.01}{$\frac{1}{3}$} } \left( \frac{m_3}{\rm TeV} \right) \left( \frac{\GeV}{m_{3/2}} \right).
\label{eq:kMin3}
\end{equation}

\section{Non-perturbative Effects}
\label{app:NonPertub}

As the sgoldstino field oscillates with a large amplitude, $v_S(T)>v_S$, the messenger field may be produced in a non-perturbative way because of the rapid change of its mass. In the main sections, we assume the non-perturbative effect is negligible, which we will now justify.

The mass of the messenger is given by
\begin{align}
m_Q^2 \simeq y^2 S^2 + g^2 T^2,
\end{align}
where $g$ is the gauge coupling constant. The adiabaticity of the mass of the messenger is characterized by the following quantity,
\begin{equation}
q\equiv \frac{|\dot{m}_Q|}{m_Q^2} \simeq \frac{y^2 |S \dot{S}|}{\left(y^2 S^2 + g^2 T^2\right)^{3/2}} .
\end{equation}

When the sgoldstino oscillates with the thermal logarithmic potential, $|\dot{S}| \simeq \alpha T^2$, and $q$ is maximized around $yS \sim gT$,
\begin{equation}
q \lsim y / (4\pi) .
\end{equation}
As long as $y< \mathcal{O}(1)$, the non-perturbative effect is negligible.

When the sgoldstino oscillates with the vacuum mass term, $|\dot{S}|\simeq m_S v_S(T)$.
We first consider the case where the sgoldstino is destroyed by scattering.
As $v_S(T) > v_S$, $S$ may vary and $q$ is maximized around $y S = gT$,
\begin{equation}
\label{eq:qmax}
q\lesssim \frac{y m_S v_S(T)}{g^2 T^2}.
\end{equation}
As long as the sgoldstino is the subdominant component of the energy density of the Universe, $q<y<1$.
After the sgoldstino dominates, $q$ grows until the thermal bath is dominated by the radiation produced from the sgoldstino at the temperature of $T_{\rm NA}$.
Using the formulae in App.~\ref{app:TdesMD}, we obtain the maximal $q$,
\begin{align}
q =  \sqrt{\frac{\pi^2 g_*}{30}}\frac{y}{g^2} D^{3/5} \lesssim D^{3/5} \frac{m_S}{m_3}~{\rm and}~10^{-2} D^{3/5} \left( \frac{m_S}{1 \, {\rm TeV}} \right)^{ \scalebox{1.01}{$\frac{1}{3}$} } ,
\end{align}
where in the inequality we use the upper bound on $y$ in Eq.~(\ref{eq:ymax}) and  $y T_{\rm des} < m_S$. We find that $q$ is smaller than unity for the parameter space considered in Fig.~\ref{fig:MasterScatt}.

We next consider the case where the sgoldstino decays. For $v_S(T) >v_S$, Eq.~(\ref{eq:qmax}) is applicable, and $q<1$ as long as the sgoldstino is subdominant.
We find that $T_S>T_M$ in the parameter space where the dilution factor is small enough that thermal leptogenesis is viable.
For the parameter region, $q<1$ for $v_S(T) > v_S$.
Once $v_S(T) < v_S$, $q$ is given by
\begin{equation}
q = \frac{y^2 m_S v_S v_S(T)}{{\rm max}(y^3 v_S^3, g^3 T^3)} < \frac{m_S}{ y v_S} \frac{v_S(T)}{v_S},
\end{equation}
which is smaller than unity as long as $m_S < y v_S$.

\end{document}